# Energetics and kinetics of vacancies in monolayer graphene boron nitride heterostructures

Bin Ouyang, Fanchao Meng and Jun Song[*]

Department of Mining and Materials Engineering,

McGill University, Montréal, Québec, H3A OC5, Canada

## Abstract

Graphene and boron nitride (GPBN) heterostructures provide a viable way to realize tunable bandgap, promising new opportunities in graphene-based nanoelectronic and optoelectronic devices. In the present study, we investigated the interplay between vacancies and graphene/*h*-BN interfaces in monolayer GPBN heterostructures. The energetics and kinetics of monovacancies and divacancies in monolayer GPBN heterostructures were examined using first-principle calculations. The interfaces were shown to be preferential locations for vacancy segregation. Meanwhile the kinetics of vacancies was found to be noticeably modified at interfaces, evidenced by the Minimum Energy Paths (MEPs) and associated migration barriers calculations. The role of interfacial bonding configurations, energy states and polarization on the formation and diffusion of vacancies were discussed. Additionally we demonstrated that it is important to recognize the dissimilarities in the diffusion prefactor for different vacancies for accurate determination of the vacancy diffusion coefficient. Our results provide essential data for the modeling of vacancies in GPBN heterostructures, and important insights towards the precise engineering of defects, interfaces and quantum domains in the design of GPBN-based devices.

---

[*] Address correspondence to be sent: jun.song2@mcgill.ca

## 1. Introduction:

The discovery of graphene along with its numerous fascinating physical and mechanical properties have resulted in a boom of research in graphene-like two dimensional (2D) nanomaterials [1-6]. Particularly, the semi-metal nature of graphene with linear dispersion at Dirac point (K) contributes to a lot of fascinating properties, e.g., extraordinary high carrier mobility and intrinsic carrier concentration, high absorbance for white light, and *etc.* [7]. All these properties make graphene a promising candidate material for a variety of nanoscale electronics and photonics devices, including high frequency transistors and high efficiency solar cell [8-11], among others. However the application of graphene is significantly limited by its characteristics of zero bandgap that inherently originated from the sublattice equivalence of carbon atoms[12]. To overcome the limitation, a variety of routes, such as functionalization[13, 14], external electric field engineering[12, 15, 16], strain[17-19]/substrate engineering[20-22] are explored to break the sublattice equivalence in order to open the bandgap of graphene. More recently, graphene-based heterostructures where graphene is integrated with other wide bandgap materials, emerge as an effective method to achieve finite bandgap while retaining essential properties of graphene [8, 23]. Among those heterostructures studied, the hybrid graphene and boron nitride (GPBN) monolayers have drawn great attention because of its unique combination of constituents: the semimetal, graphene, married with wide band-gap (i.e., 4.6eV) insulator, hexagonal boron nitride (*h*-BN). Graphene and *h*-BN have the same honeycomb structure with similar lattice constants (i.e., ~1.6% lattice mismatch), contributing to little distortion along the coherent interface. The GPBN heterostructure also possesses large space of domain size dominated band gap engineering which contribute to high on/off ratios for nanoelectronic devices[8].

The monolayer GPBN heterostructures are commonly fabricated using chemical vapour deposition (CVD)[23, 24]. Recently it is shown by Yong Ji et al.[25] that topological substitution reaction augmented by lithography provides another route to fabricate the GPBN heterostructure with more precise control of the domain shape and size. In general, the fabrication process of GPBN heterostructures involves high temperature and substrates with dissimilar lattices as the host material, thus producing abundant defects. One prevailing category of defects in GPBN heterostructures are vacancies. It has been shown by Refs [19, 26] that vacancies can noticeably affect the electronic properties of graphene and *h*-BN. In particular for the monolayer GPBN heterostructure, vacancies may interact with its inherent structural heterogeneities, i.e., graphene/*h*-BN interfaces, to modify the interface structure and subsequently the domain size and geometry. Therefore it is important to understand the thermodynamics and kinetics of vacancies in the GPBN heterostructure. However, to date comprehensive knowledge about vacancy segregation and migration at interfaces in GPBN heterostructures remains largely absent, although the existence and importance of vacancies in GPBN heterostructures are well recognized [27-29].

In the present work, we systematically investigate the energetics and kinetics of both mono- and divacancies in monolayer GPBN heterostructures where graphene and *h*-BN are connected via either zigzag or armchair interface, using first-principle calculations. The remainder of the paper is organized as follows. Section 2 describes the computational methodology used in our study. In Section 3, the computed formation energies, and migration paths and barriers of vacancies are presented, following which the interplay between vacancies and graphene/*h*-BN interfaces is discussed. In addition, preliminary calculations of the jumping frequencies associated with vacancy diffusion in the GPBN heterostructure are also presented. Finally we summarize our results and discuss their implications to defect evolution and engineering in GPBN heterostructures in Section 4.

## 2. Methodology:

Density functional theory (DFT)[30] calculations are performed using the Vienna Ab-initio Simulation Package (VASP)[31]. The projector augmented-wave (PAW) potentials has been adopted while the generalized gradient approximation (GGA) method with functional developed by Perdew, Burke and Ernzerhof (PBE) has been used[32]. Due to the existence of dangling bonds in vacancy decorated system, spin polarization is considered in all calculations. In addition, climbed image nudged elastic band (ci-NEB)[33] is employed to identify the minimum energy paths and transition states during vacancy migration.

The GPBN heterostructure is constructed by welding nanoribbons of graphene and *h*-BN together. The graphene and *h*-BN are of equal molar fraction and are connected with each other via either zigzag (ZZ) or armchair (AC) interfaces, as illustrated in Figure 1. Particularly we note that there are two types of ZZ interfaces, one with C-N bonds (see Fig. 1a) and the other with C-B bonds (see Fig. 1b) across the interface. Simulation cells of cell dimensions 6× 16 (192 atoms) and 6× 12 (144 atoms) are constructed for GPBN heterostructures with AC interfaces (denoted as AC-GPBN below) and ZZ interfaces (denote as ZZ-GPBN below) respectively. In the ZZ-GPBN heterostructure, both types of ZZ interfaces (i.e., with C-B or C-N bonds) are present. Benchmark studies have been performed to ensure that the cell dimensions chosen are large enough to eliminate interactions between defects and interface as well as their periodic images. A *k*-point grid of $5\times7\times1$ and energy cutoff of 600 eV are used in the DFT calculations. The lattice constant of the heterostructure is set as 2.49 Å which is proven to yield the lowest energy for the system.

Vacancies are introduced into the GPBN heterostructure by removing individual atoms (for monovacancies) or atom duals (for divacancies). In particular, the creation of different vacancies along graphene/*h*-BN interfaces is outlined in Table I with relevant atom sites indicated in Figure 1. In the following context we denote the monovacancy as $SV_\alpha$ and divacancy as $DV_{\beta\gamma}$ where $\alpha =$ C, B or N and $\beta\gamma =$ CC, CB or BN denote the corresponding species of individual atoms and atom duals removed during vacancy creation, to indicate the types for $SV_\alpha$ and $DV_{\beta\gamma}$ respectively. The formation energies $E_f$s of $SV_\alpha$ and $DV_{\beta\gamma}$ are defined as follows:

$$E_f(SV_\alpha) = E_{tot}^V(SV_\alpha) - E_{tot}^{GPBN} + \mu_\alpha ,$$
$$E_f(DV_{\beta\gamma}) = E_{tot}^V(DV_{\beta\gamma}) - E_{tot}^{GPBN} + \mu_\beta + \mu_\gamma ,$$
(1)

where in this formula, the $E_{tot}^{GPBN}$ and $E_{tot}^V$ are the total energies of the system before and after the introduction of vacancy, and $\mu_i$ ($i =$ C, B or N) is the chemical potential of the corresponding atom *i* removed during vacancy creation[34, 35]. The above formula is introduced by Laaksonen[36] and applies when we are dealing with the charge-neutral defects in graphene like structures[27, 29].

## 3. Results and Discussion:

The ground state atomic configurations and local charge transfer contours of different monovacancies are shown in Figure 2. Table II lists the formation energies of monovacancies in graphene, *h*-BN, ZZ-GPBN and AC-GPBN. As shown in the table, our results for graphene and *h*-BN are consistent with previous studies in literature[26, 35]. Meanwhile we see from the table that the $E_f$ for an interface $SV_\alpha$ is substantially lower than the corresponding bulk one, suggesting a strong tendency for monovacancies to segregate towards the interface. In addition, we see that the ZZ interface is energetically preferred over the AC interface, as evidenced by the lower $E_f$ values. In particular for $SV_C$, there are two possible configurations, *i.e.*, the vacancy neighboring a B atom (cf. Figs. 2c and 2g) or neighboring an N atom (cf. Figs 2d and 2h), at both ZZ and AC interfaces. It is also observed that $SV_C$ exhibits lower $E_f$ when neighboring the B atom, which is likely due to the C-B bond being weaker than the C-N bond [26, 35, 37, 38].

The lower formation energies of monovacancies at interfaces mainly comes from the local high energy states that renders the removal of atoms easier. The elevation of energy states at the interface originated from the C-B and C-N bonds that induce fluctuations in local potentials for GPBN heterostructures (See Supplement Information). Furthermore we note that the energy states at ZZ and AC interfaces are different, with the interface formation energy being 0.26 eV/Å for the ZZ interface and 0.17 eV/Å unit cell for the AC interface. This is consistent with the fact that $SV_\alpha$ exhibits lower $E_f$ values at the ZZ interface than the AC interface. The higher interface formation energy of the ZZ interface possibly derives from the large polarization at the ZZ interface, as illustrated by the charge transfer plots in Figs 2-3.

As monovacancies aggregate at the interface, they may coalescence into divacancies or even vacancy clusters. In this regard, we examine the formation of divacancies that represent the next step in the evolution of vacancies. The ground state atomic configurations and local charge difference contours of different divacancies are shown in Figure 3. The formation energies of divacancies in graphene, $h$-BN, AC-GPBN and ZZ-GPBN are shown in Table III. Several observations can be drawn from the data. Firstly we see that the $E_f$ values exhibit a trend of $E_f$ ($DV_{CC}$) < $E_f$ ($DV_{CN}$) < $E_f$ ($DV_{CB}$) < $E_f$ ($DV_{BN}$). This trend is consistent with the energetics of monovacancies shown in Table II if we consider a divacancy $DV_{\beta\gamma}$ as a combination of two monovacancies, i.e., $SV_\beta$ and $SV_\gamma$. On the other hand, the $E_f$ of a divacancy is smaller than the net summation of the $E_f$ values of the two corresponding monovacancies, suggesting that the coalescence of monovacancies into divacancies is energetically favorable. The energy release accompanying the coalescence of monovacancies into divacancies is probably due to the local bond reconstruction, as evidenced by the overlapping of charge clouds around divacancies demonstrated in Figure 3, as also noted in previous studies of divacancies in graphene and *h*-BN[26, 35]. Secondly we note that, similar to the case of monovacancies (cf. Table II), divacancies also show substantially lower formation energies at interfaces than in the bulk (i.e., graphene or *h*-BN) and a preference of the ZZ interface over AC interface, which again can be attributed to the high energy states at interfaces and different interface formation energies as previously discussed. In addition we note that $DV_{CC}$ and $DV_{BN}$ each exhibit two $E_f$ values corresponding to the two types of ZZ interfaces (cf. Figs 1a-b), with $E_f$ being slightly lower in the interface with C-N bonds.

### 3. 1. Kinetics of vacancies:

With the energy states of vacancies at interfaces being quite different from those in bulk (i.e., pristine graphene or *h*-BN), the migration kinetics of vacancies is also expected to be modified at interfaces. The non-identical diffusion paths of vacancies at different interfaces are illustrated in Figure 4. For monovacancies, the migration occurs through the vacancy exchanging with one of its neighboring atoms. In particular, for $SV_C$ the diffusion occur by the swap of the vacancy with any of its immediate neighboring atoms (cf. Figs. 4b and d), while for $SV_B$ (or $SV_N$) the diffusion occur by the swap of the vacancy with its neighboring C atom or B (or N) atom (cf. Figs. 4a and 4c). One thing worth noting is that the swap between $SV_B$ (or $SV_N$) and N (or B) atom does not happen as it would result in the N-N (or B-B) homo-elemental bond that is rendered unstable by the large coulomb repulsion.

Using NEB calculations, the Minimum Energy Paths (MEPs) are computed for different monovacancies. Three representing MEPs are presented in Figure 5, showing that the MEP may consist of either single barrier or double barriers. The double-barrier MEP stems from the binary nature of the *h*-BN phase and the symmetry associated with the B-N bond. The energy barrier ($E_b$) values for different migration paths are listed in Table IV. In addition, the migration barriers for monovacancies in graphene (i.e., 1.3 eV for $SV_C$ [26]) and *h*-BN (i.e., 3.3 eV for $SV_B$ and 6.1 eV for $SV_N$ [35]) are also listed.

Examining the migration paths of monvacancies in Figure 4, we note that in general the motion direction is either along (or close to being along) or perpendicular to the interface. The two motion directions are indicated by ∥ and ⊥ symbols in Table IV for simplicity. In addition, we can see that for the ∥ migration motion the vacancy type (i.e., the value of $\alpha$ in $SV_\alpha$) remains the same while for the ⊥ migration motion the vacancy type may alter. In this regard, we add a superscript "*a*" to the ⊥ symbol (cf. Fig. 4 and Table IV) to separately denote the motion that results in a type change. There are several trends we can note from Table IV. Firstly the $E_b$ value is always lower in the ZZ interface compared to the corresponding one in the AC interface, likely due to the polarization at the ZZ interface that facilitates the bond breaking/forming process during the vacancy migration. Secondly the vacancies exhibit much different migration barriers along different motion directions. In particular for the ∥ motion, the migration of $SV_C$ is inhibited while the migrations of $SV_B$ and $SV_N$ are facilitated at the interface, compared to the bulk. This can be understood from the strength of bonding in the GPBN, which exhibits a trend of C-C < C-B < C-N < B-N as shown by the previous research [26, 35, 37, 38]. Consequently the bonding that resists the ∥ motion of $SV_C$ at the interface is strengthened while it is weakened for the cases of $SV_B$ and $SV_N$. Thirdly for the ⊥ motion at the interface, the $E_b$ value for each vacancy is slightly higher than the one in the bulk. This is expected as the ⊥ motion results in the vacancy moving from the interface to the bulk, a higher-energy location for the vacancy. On the other hand, no clear trend in $E_b$ is observed for the $\perp^a$ motion. Another thing worth mentioning is that for the N atom in either ZZ or AC interface, our calculations show that the $\perp^a$ motion is not viable though geometrically possible.

For divacancies at the interface, they are found to migrate via rotation which yields the lowest energy barrier, consistent with previous studies [26, 35, 38]. The rotation motion occurs by an atom immediately neighboring the divacancy swapping with one of the two missing atoms sites that constitute the divacancy, as illustrated in Figure 4 where the possible rotation directions for each vacancy are numbered and indicated by arrows. The MEPs for the migration of those divacancies are computed, with three representing MEP curves shown in Figure 5. The corresponding $E_b$ values are listed in Table V. One thing worth noting is that the divacancy may also migrate by dissociation/recombination of two single vacancies. In this regard, we did some preliminary studies and found that the barrier associated with the dissociation/recombination is much higher than the one for the rotation motion. As a consequence, the paths/barriers listed in Table V are appropriate for describing the predominating migration behaviors of divacancies.

From Table V, we can note that the $E_b$ values for the divacancy diffusion at ZZ and AC interfaces in GPBN range from (approximately) 4.0 to 7.5 eV, overall being in the same ballpark as the ones for bulk divacancies despite larger variation. Viewing a divacancy as the coalescence of two monvacancies, we note that in general for an interface divacancy, its $E_b$ is much higher than the $E_b$ values of the two corresponding monovacancies (cf. Table IV). This suggests that divacancies are much less mobile than monovacancies at the interface in the GPBN. We also note from Table V that the diffusion of $DV_{CC}$ is facilitated while the diffusion of $DV_{BN}$ is moderated at interfaces compared to the graphene and $h$-BN bulk phases. To understand this phenomenon, we first examined the diffusion of $DV_{CC}$ and $DV_{BN}$ in their bulk phases. From Table V we see that $E_b$ of DV$_{CC}$ in graphene is about 7 eV, higher than the $E_b$ values of DV$_{BN}$ in *h*-BN are 6.0 eV and 4.5 eV corresponding to rotation motions with B moving and N moving respectively, suggesting that the C-C bond is more resistant to the rotation motion of divacancy than the B-N bond. These two bonding environments mesh with each other at the interface, and consequently we expect the resistance to the rotation of divacancy to stay in-between, i.e., the resistance (barrier) for the motion of DV$_{CC}$ would decrease while the one for DV$_{BN}$ would increase, consistent with the observation. Another observation we can draw from the data in Table V is that the $E_b$ of DV$_{CB}$ or DV$_{CN}$ is in general lower in the ZZ interface than in the AC interface, similar to the case of monovacancies (cf. Table IV). This is likely also attributed to the polarization at the ZZ interface as previously discussed.

### 3. 2. Diffusion of vacancies along the interface:

The energy barrier, $E_b$, provides crucial information to understand the migration kinetics of vacancies in GPBN, evidenced by the equation below:

$$D = D_0 \exp(-\frac{E_a}{k_B T}) = ga^2 \tilde{v} \exp(-\frac{E_a}{k_B T}), \tag{1}$$

where $D$ denotes the diffusivity of the vacancy migration in GPBN, $g$ denotes the geometry constant, $a$ denotes the distance of each hoping, $E_a$ indicates the migration barrier, $\tilde{v}$ denotes the effective jumping frequency, $D_0 = ga^2\tilde{v}$ is the diffusion prefactor, while $k_B$ and $T$ are the Boltzmann constant and temperature respectively. From Eq. 1 above, we see that besides $E_b$, information of the temperature-independent prefactor $D_0$, is also necessary to completely prescribe $D$. In the prefacor $D_0$, $g$ is a geometrical constant derivable from the lattice geometry for atom (vacancy) hopping[39](often approximated as unity for 2D material systems[40]) and $a$ can be directly obtained given the migration path. The parameter $\tilde{v}$, on the other hand, can be derived from the lattice vibrations at the initial and saddle point states for each jump based on the transition states theory[41]. Nonetheless, to our knowledge there has yet been any study directly computing $\tilde{v}$ for 2D material systems. Often $\tilde{v}$ is simply approximated using the Debye frequency or some other estimated constants [42, 43]. The simple approximation however may be inaccurate as shown by Toyoura et al.[41] In this regard, we performed some preliminary studies to directly evaluate $\tilde{v}$ by computing the eigenfrequencies from first-principle calculations in order to enhance the accuracy in the prediction of $D_0$, elaborated below.

According to Vineyard[44], $\tilde{v}$ can be evaluated according to the following equation:

$$\tilde{v} = \prod_{i=1}^{N} v_i^I \Big/ \prod_{i=1}^{N-1} v_i^S , \qquad (2)$$

where $v_i^I$ and $v_i^S$ are the frequencies of the normal vibration modes at the initial and saddle points respectively. Using Eqs. 1-2 above together with the DFT calculations, the values of $\tilde{v}$ and $D_0$ are obtained for $SV_C$ in graphene, and $SV_B$ and $SV_N$ in $h$-BN, listed in Table VI. The computed $\tilde{v}$ values are also compared with the estimates (cf. Table V) using the Debye model[45] given as,

$$v_m = (3N/4\pi V)^{1/3} v_s , \qquad (3)$$

with $V$ and $v_s$ being the corresponding volume[46] and speed of sound respectively. We can see that though the two set of frequency data are close in values and exhibit a similar trend, the one computed from first-principle calculations recognizes the difference between $SV_B$ and $SV_N$ while the Debye model does not. The $\tilde{v}$ and $D_0$ data, along with the $E_b$ data previously obtained, provide essential inputs for determining the diffusivities of vacancies in GPBN heterostructures [†].

---

[†] Please note that though the vacancies at interfaces in the GPBN heterostructure would have different $\tilde{v}$ and $D_0$ values (the precise determination of $\tilde{v}$ and $D_0$ would require a separate set of DFT calculations for each vacancy configuration/MEP path), the data in Table VI provides a first-order approximation. For instance, the $D_0$ for a $SV_\alpha$ at

## 4. Conclusion:

To summarize, the energetics and kinetics of vacancies at zigzag (ZZ) and armchair (AC) interfaces in monolayer graphene and boron nitride (GPBN) heterostructures were examined using first-principle calculations. Our results show that interfaces in GPBN heterostructures provide energetically favorable locations for vacancy segregation. The preferential segregation of vacancies at interfaces was shown to be directly related to the high energy states at interfaces that ease the formation of vacancies. The exact energetics of vacancies were found to be dependent on local bonding geometry and polarization. In addition, our results indicate that the coalescence of monovacancies into divacancies is energetically preferred at the interface, thus being potentially the next step in vacancy evolution following the segregation.

The Minimum Energy Paths (MEPs) and associated migration barriers were obtained for vacancies at the ZZ and AC interfaces, showing noticeable effects of interfaces on vacancy migration. For monovacancies, we found that the migration motions along interfaces are inhibited for the C vacancy, but facilitated for the B and N vacancies, in comparison to corresponding bulk phases (i.e., graphene and $h$-BN), which was attributed to the characteristic bonding configurations at the interface. On the other hand, for the migration motions of monovacancies perpendicular to interfaces, we found the migration barrier becomes higher compared to the bulk diffusion if the motion results in the vacancy escaping the interface into the bulk while the migration barrier varies if the motion results in the vacancy moving across the interface. The divacancies in GPBN heterostructures were found to migrate via rotation motions with their migration barriers being much higher than those of monovacancies, both at the interface and in the bulk. One particular

---

the interface can be approximated by the $D_0$ value for a $SV_\alpha$ in the bulk, while the $D_0$ for a divacancy $DV_{\alpha\beta}$ via path $\hat{\alpha}\beta(\alpha\hat{\beta}) \xrightarrow{\text{Path}} \underline{\delta}\beta(\alpha\underline{\delta})$ can be approximated by the $D_0$ value for a $SV_\alpha$ ($SV_\beta$) in the bulk.

observation was that the migration of $DV_{CC}$ is facilitated while the migration of $DV_{BN}$ is moderated at interfaces, compared to corresponding bulk phases. Besides the migration barriers, preliminary calculations of the jumping frequencies associated with vacancy diffusion in the GPBN heterostructure are presented, showing that first-principle calculations offer a means to directly compute the diffusion prefactor. It was further demonstrated that the first-principle approach can effectively recognize the dissimilarities in the diffusion prefactor for different vacancies whilst the often-used Debye model does not.

The present study clarifies the interactions between vacancies and graphene/*h*-BN interfaces and provides essential data for modeling vacancy nucleation, diffusion and coalescence in GPBN heterostructures. Furthermore, with the strong preferential segregation of vacancies at graphene/*h*-BN interfaces, our findings provide direct insights towards interface engineering in GPBN heterostructures, thus of great relevance to the precise manipulation of quantum domains and subsequently the material properties in the design of GPBN-based devices.

**Table I**: The list of interface vacancies considered, with the numbers indicating the corresponding sites of atoms/atom duals removed from the GPBN heterostructure to generate mono- and divacancies.

| Monovacancy | Atoms removed | Divacancy | Atom duals removed |
| --- | --- | --- | --- |
| $SV_C$ | I, V, IX, XI | $DV_{CC}$ | (I, III), (V, VII), (IX, XI) |
| $SV_B$ | II, X | $DV_{CB}$ | (I, II), (IX, X) |
| $SV_N$ | VI, XII | $DV_{CN}$ | (V, VI), (XI, XII) |
| | | $DV_{BN}$ | (II, IV), (VI, VIII), (XII, XIII) |

**Table II**. Formation energies of different monovacancies in graphene, *h*-BN, ZZ-GPBN and AC-GPBN. For the SV$_C$, there are two $E_f$ values separated by a slash, with the first and second corresponding to the configurations of the vacancy neighboring a B atom and an N atom respectively. The numbers in parentheses are $E_f$ data taken from literature.

| Material system | $E_f$ (eV) | | |
|---|---|---|---|
| | SV$_C$ | SV$_B$ | SV$_N$ |
| Graphene | 7.50 (7.57)[26] | N/A | N/A |
| Boron Nitride | N/A | 10.88 (11.22)[35] | 8.81 (8.91)[35] |
| ZZ-GPBN | 5.60/6.08 | 8.29 | 7.23 |
| AC-GPBN | 6.18/7.01 | 8.83 | 8.34 |

**Table III**. Formation energies of different divacancies in graphene, *h*-BN, ZZ-GPBN and AC-GPBN. For the $DV_{CC}$ and $DV_{BN}$ in the ZZ-GPBN heterostructure, there are two $E_f$ values, with the first and second corresponding to ZZ interfaces with C-B bonds (cf. Fig. 1a) and C-N bonds (cf. Fig. 1b) respectively. The numbers in parentheses are $E_f$ data taken from literature.

| System | $DV_{CC}$ (eV) | $DV_{CB}$ (eV) | $DV_{CN}$ (eV) | $DV_{BN}$ (eV) |
|---|---|---|---|---|
| Graphene | 7.47(7-8)[26] | N/A | N/A | N/A |
| Boron Nitride | N/A | N/A | N/A | 11.91(11.73)[35] |
| ZZ-GPBN | 5.36/5.35 | 10.25 | 8.41 | 10.32/10.17 |
| AC GPBN | 6.23 | 10.35 | 8.75 | 10.83 |

**Table IV:** Migration paths and corresponding energy barriers, $E_b$ s for different monovacancies at ZZ and AC interfaces. The superscripts, B and N, in $SV_C$ indicate the two different C monovacancies, at ZZ interfaces with C-B bonds (cf. Fig. 1a) and C-N bonds (cf. Fig. 1b) respectively. Different vacancy migration paths at the interface are indicated by the symbols ∥, ⊥ and $⊥^a$ as described in the text. The $\alpha \rightarrow \beta$ describes the evolution of vacancy type during the migration, where the vacancy constituent $\alpha$ swaps with an atom of specie $\beta$. The migration data (from literature) of $SV_C$ in graphene, and $SV_B$ and $SV_N$ in $h$-BN are also listed.

| | \multicolumn{12}{c}{**ZZ Interface**} |
|---|---|---|---|---|---|---|---|---|---|---|---|---|
| Type | $SV_C^B$ | | | $SV_C^N$ | | | $SV_B$ | | | $SV_N$ | | |
| Path | ∥ | | $⊥^a$ | ∥ | | $⊥^a$ | ∥ | ⊥ | $⊥^a$ | ∥ | ⊥ | $⊥^a$ |
| $\alpha \rightarrow \beta$ | C→C | | C→B | C→C | | C→N | B→B | B→B | B→C | N→N | N→N | N→C |
| $E_b$ (eV) | 2.17 | | 0.95 | 3.09 | | 2.15 | 2.27 | 3.45 | 2.10 | 4.03 | 6.20 | N/A |
| | \multicolumn{12}{c}{**AC Interface**} |
| Type | $SV_C^B$ | | | $SV_C^N$ | | | $SV_B$ | | | $SV_N$ | | |
| Path | ∥ | ⊥ | $⊥^a$ | ∥ | ⊥ | $⊥^a$ | ∥ | ⊥ | $⊥^a$ | ∥ | ⊥ | $⊥^a$ |
| $\alpha \rightarrow \beta$ | C→C | C→C | C→B | C→C | C→C | C→N | B→B | B→B | B→C | N→N | N→N | N→C |
| $E_b$ (eV) | 2.81 | 1.45 | 1.14 | 3.24 | 1.48 | 3.31 | 2.83 | 3.30 | 2.68 | 4.66 | 6.02 | N/A |

| **Graphene** | **$h$-BN** | |
|---|---|---|
| $SV_C$: $E_b$ = 1.3 eV[26] | $SV_B$: $E_b$ = 2.6 eV[35] | $SV_N$: $E_b$ = 5.8 eV[35] |

**Table V**: Migration paths and corresponding energy barriers, $E_b$ s for different divacancies at ZZ and AC interfaces. The superscripts, B and N, in $DV_{CC}$ indicate the two C divacancies, at ZZ interfaces with C-B bonds (cf. Fig. 1a) or C-N bonds (cf. Fig. 1b) respectively. The symbols ①, ②, ③ and ④ denote different vacancy migration paths at the interface as illustrated in Figure 4. The $\hat{\alpha}\beta(\alpha\hat{\beta}) \xrightarrow{\text{Path}} \underline{\delta}\beta(\alpha\underline{\delta})$ describes the evolution off vacancy type following a particular migration path, where the vacancy constituent, indicated by the capped symbol (i.e., $\hat{\alpha}$ or $\hat{\beta}$) on the left side, rotates to swap with the atom, indicated by the underlined symbol (i.e., $\underline{\delta}$) on the right side. The migration data (from literature) of $DV_{CC}$ in graphene, and $DV_{BN}$ in $h$-BN are also listed.

| ZZ Interface | | | | | | | | |
|---|---|---|---|---|---|---|---|---|
| Type | $DV_{CC}^{B}$ | | | | $DV_{CC}^{N}$ | | | |
| $\hat{\alpha}\beta(\alpha\hat{\beta}) \xrightarrow{\text{Path}} \underline{\delta}\beta(\alpha\underline{\delta})$ | $\hat{C}C \xrightarrow{①} \underline{C}C$ | $C\hat{C} \xrightarrow{②} C\underline{C}$ | $C\hat{C} \xrightarrow{③} C\underline{C}$ | $\hat{C}C \xrightarrow{④} \underline{B}C$ | $\hat{C}C \xrightarrow{①} \underline{C}C$ | $C\hat{C} \xrightarrow{②} C\underline{C}$ | $C\hat{C} \xrightarrow{③} C\underline{C}$ | $\hat{C}C \xrightarrow{④} \underline{N}C$ |
| $E_b$ (eV) | 6.11 | 5.95 | 5.68 | 5.17 | 6.08 | 6.18 | 5.67 | 3.98 |
| Type | $DV_{BN}^{B}$ | | | | $DV_{BN}^{N}$ | | | |
| $\hat{\alpha}\beta(\alpha\hat{\beta}) \xrightarrow{\text{Path}} \underline{\delta}\beta(\alpha\underline{\delta})$ | $\hat{B}N \xrightarrow{①} \underline{B}N$ | $B\hat{N} \xrightarrow{②} B\underline{N}$ | $\hat{B}N \xrightarrow{③} \underline{B}N$ | $B\hat{N} \xrightarrow{④} B\underline{C}$ | $B\hat{N} \xrightarrow{①} B\underline{N}$ | $\hat{B}N \xrightarrow{②} \underline{B}N$ | $B\hat{N} \xrightarrow{③} B\underline{N}$ | $\hat{B}N \xrightarrow{④} \underline{C}N$ |
| $E_b$ (eV) | 4.91 | 6.04 | 4.79 | 5.60 | 4.92 | 6.11 | 4.84 | 7.17 |
| Type | $DV_{CB}^{B}$ | | | | $DV_{CN}^{N}$ | | | |
| $\hat{\alpha}\beta(\alpha\hat{\beta}) \xrightarrow{\text{Path}} \underline{\delta}\beta(\alpha\underline{\delta})$ | $C\hat{B} \xrightarrow{①} C\underline{C}$ | $\hat{C}B \xrightarrow{②} \underline{N}B$ | | | $C\hat{N} \xrightarrow{①} C\underline{C}$ | $\hat{C}N \xrightarrow{②} \underline{B}N$ | | |
| $E_b$ (eV) | 4.46 | 4.27 | | | 5.51 | 5.60 | | |
| AC Interface | | | | | | | | |
| Type | $DV_{CC}^{*}$ | | | | $DV_{BN}$ | | | |
| $\hat{\alpha}\beta(\alpha\hat{\beta}) \xrightarrow{\text{Path}} \underline{\delta}\beta(\alpha\underline{\delta})$ | $\hat{C}C \xrightarrow{①} \underline{B}C$ | $C\hat{C} \xrightarrow{②} C\underline{N}$ | $\hat{C}C \xrightarrow{③^B} \underline{C}C$ | $C\hat{C} \xrightarrow{③^N} C\underline{C}$ | $B\hat{N} \xrightarrow{①} B\underline{N}$ | $\hat{B}N \xrightarrow{②} \underline{B}N$ | $B\hat{N} \xrightarrow{③} \underline{C}B$ | $\hat{B}N \xrightarrow{④} \underline{C}N$ |
| $E_b$ (eV) | 4.45 | 4.04 | 6.68 | 6.80 | 5.00 | 6.21 | 6.65 | 7.53 |
| Type | $DV_{CB}$ | | | | $DV_{CN}$ | | | |
| $\hat{\alpha}\beta(\alpha\hat{\beta}) \xrightarrow{\text{Path}} \underline{\delta}\beta(\alpha\underline{\delta})$ | $C\hat{B} \xrightarrow{①} C\underline{C}$ | $C\hat{B} \xrightarrow{②} B\underline{N}$ | $C\hat{B} \xrightarrow{③} C\underline{C}$ | $\hat{C}B \xrightarrow{④} B\underline{N}$ | $C\hat{N} \xrightarrow{①} C\underline{C}$ | $\hat{C}N \xrightarrow{②} \underline{B}N$ | $C\hat{N} \xrightarrow{③} C\underline{C}$ | $\hat{C}N \xrightarrow{④} \underline{B}N$ |
| $E_b$ (eV) | 6.07 | 5.32 | 5.92 | 5.41 | 4.71 | 6.56 | 4.45 | 6.62 |

| Graphene | $h$-BN | |
|---|---|---|
| $E_b$ = 7.0 eV[26] | Rotation via B moving $E_b$ = 6.0 eV[35] | Rotation via N moving $E_b$ = 4.5 eV[35] |

**Table VI**: The effective jumping frequency values, computed directly from DFT calculations (i.e., $\tilde{v}$) and estimated from the Debye model (i.e., $v_m$, see Eq. 3), for $SV_C$ in graphene, and $SV_B$ and $SV_N$ in $h$-BN. The corresponding values of the prefactor $D_0 = ga^2\tilde{v}$ are also listed assuming $g = 1$.

| Material system | Graphene | $h$-BN | |
|---|---|---|---|
| Atom type | C | B | N |
| $\tilde{v}$ (THz) | 9.84 | 9.24 | 8.67 |
| $v_m$ (THz) | 11.93 | 26.04 | 26.04 |
| $D_0 = ga^2\tilde{v}$ (m²/s) | $5.95 \times 10^{-7}$ | $5.87 \times 10^{-7}$ | $5.51 \times 10^{-7}$ |

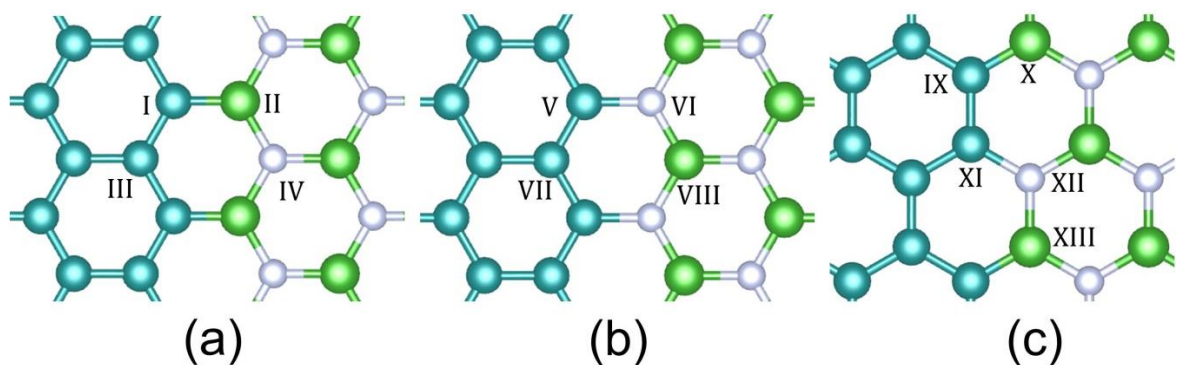

**Figure 1**: Local atomic configurations of the ZZ interfaces with *a*) C-B bonds and *b*) C-N bonds respectively, and *c*) the AC interface, in the GPBN heterostructures. The C, B and N atoms are colored dark cyan, green and whiter respectively. The numbered atoms along interfaces indicate the sites considered for vacancy creation (cf. Table I).

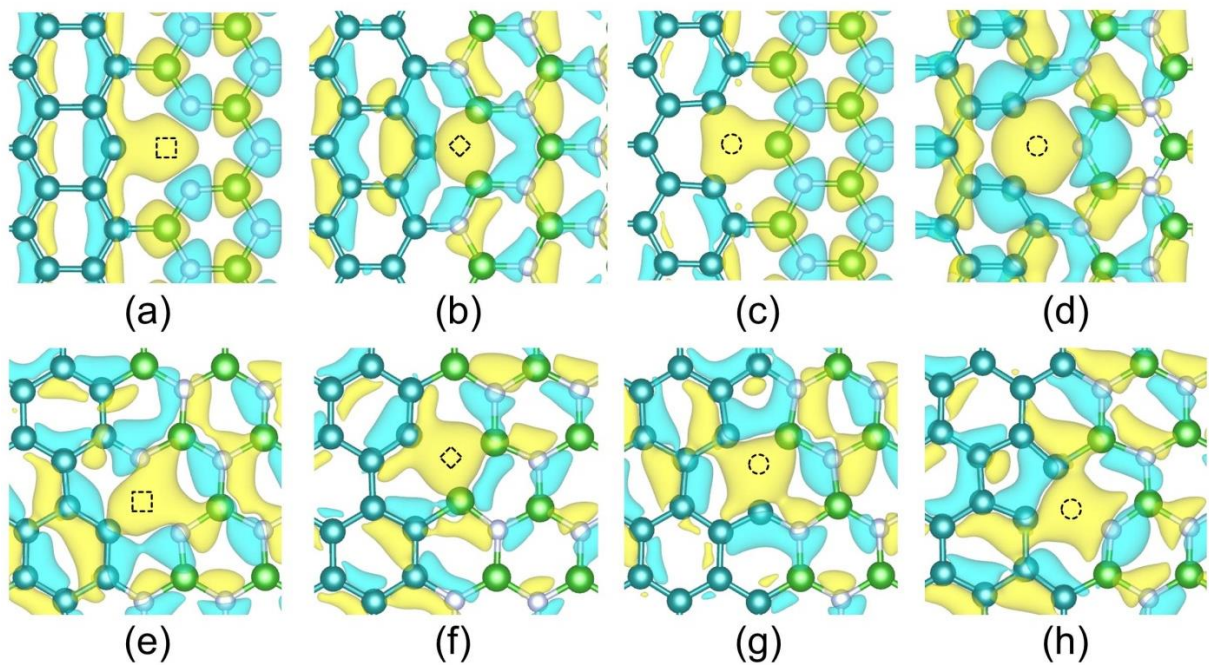

**Figure 2:** Ground states configurations and charge transfer contours of monovacancies: *a)* SV$_B$, *b)* SV$_N$, *c)* SV$_C$ immediately neighboring B, and *d)* SV$_C$ immediately neighboring N in ZZ interface, and *a)* SV$_B$, *b)* SV$_N$, *c)* SV$_C$ immediately neighboring B, and *d)* SV$_C$ immediately neighboring N in AC interface. The C, B and N atoms are colored dark cyan, blue and white respectively. The dashed circle, square and diamond symbols indicate the C, B and N atoms respectively, that are removed during the creation of monovacancies.

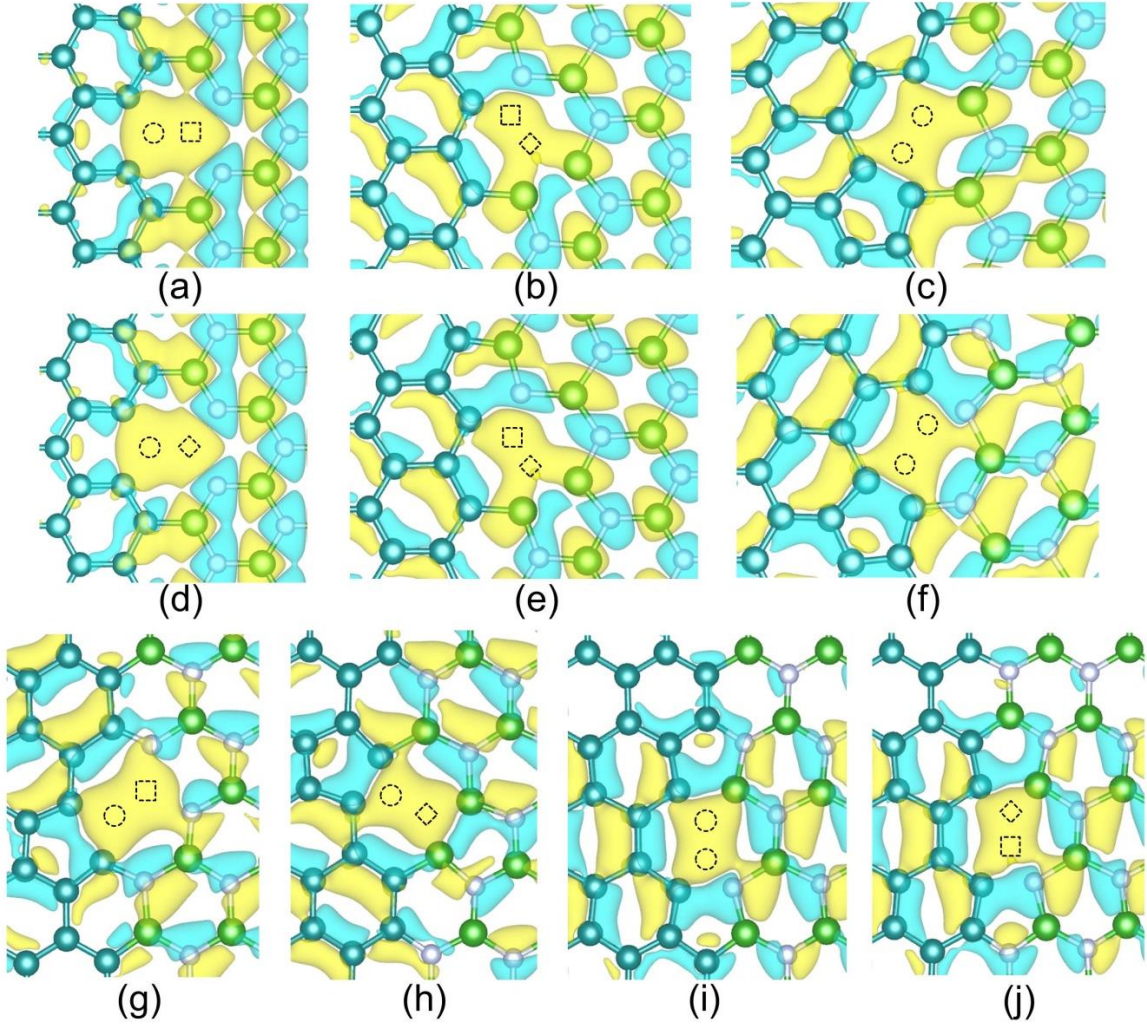

**Figure 3:** Ground states and charge transfer contours of divacancies: *a*) $DV_{CB}$, *b*) $DV_{BN}$ and *c*) $DV_{CC}$ at the ZZ interface with C-B bonds, *d*) $DV_{CN}$, *e*) $DV_{BN}$ and *f*) $DV_{CC}$ at the ZZ interface with C-N bonds, and *g*) $DV_{CB}$, *h*) $DV_{CN}$, *i*) $DV_{CC}$ and *j*) $DV_{BN}$ at the AC interface. The C, B and N atoms are colored dark cyan, blue and white respectively. The dashed circle, square and diamond symbols indicate the C, B and N atoms respectively, that are removed during the creation of divacancies.

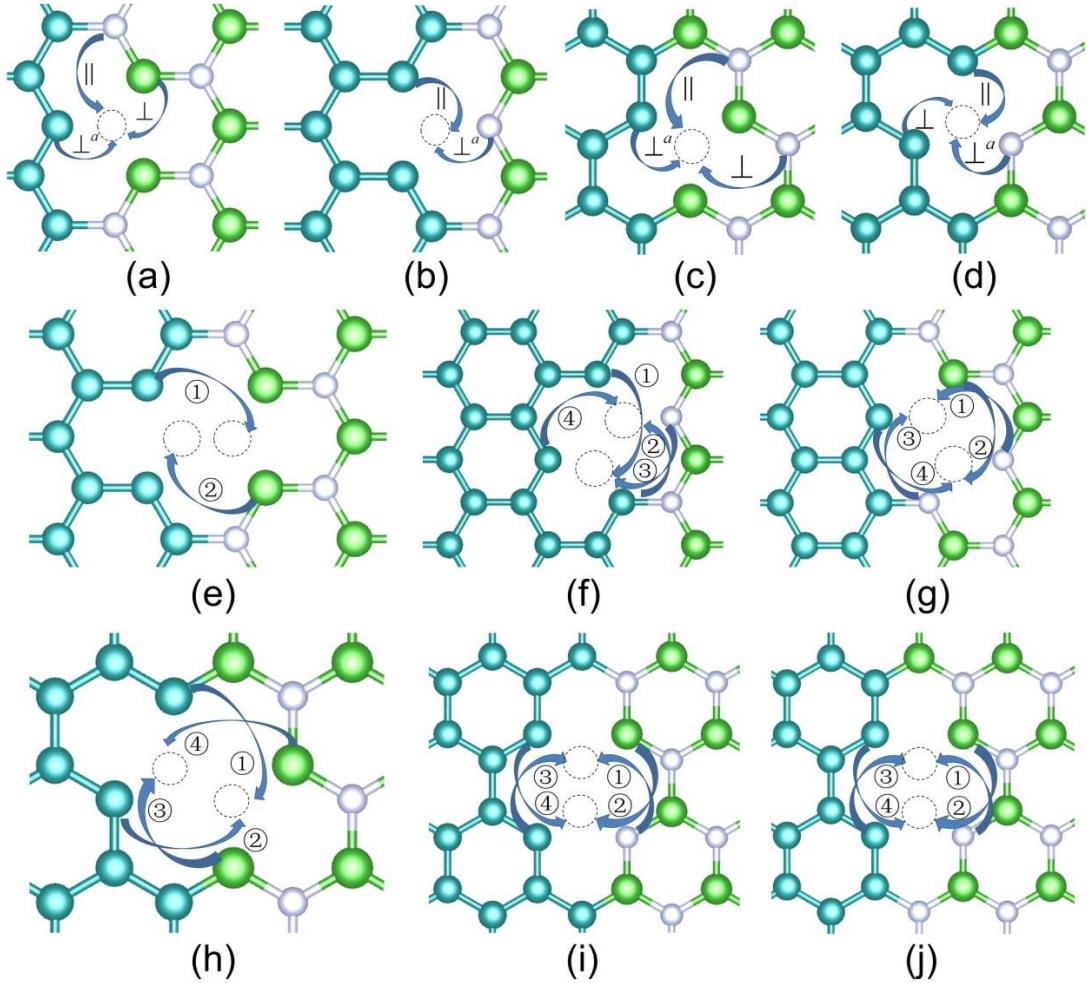

**Figure 4:** Possible migration paths of monovacancies: *a)* $SV_N$ (or $SV_B$) and *b)* $SV_C$ at the ZZ interface, and *c)* $SV_B$ (or $SV_N$) and *d)* $SV_C$ at the AC interface, and divacancies: *e)* $DV_{CN}$ (or $DV_{CB}$), *f)* $DV_{CC}$ and *g)* $DV_{BN}$ at the ZZ interface, and *e)* $DV_{CN}$ (or $DV_{CB}$), *f)* $DV_{CC}$ and *g)* $DV_{BN}$ at the AC interface. The symbols $\parallel$, $\perp$ and $\perp^a$ denote possible migration paths for monovacancies, and the symbols ①, ②, ③ and ④ denote possible migration paths for divacancies at the interface, as described in the text. The green and white atoms represent the B and N atoms (interchangeable due to the symmetry of the *h*-BN phase), while the dark cyan atoms represents the C atoms.

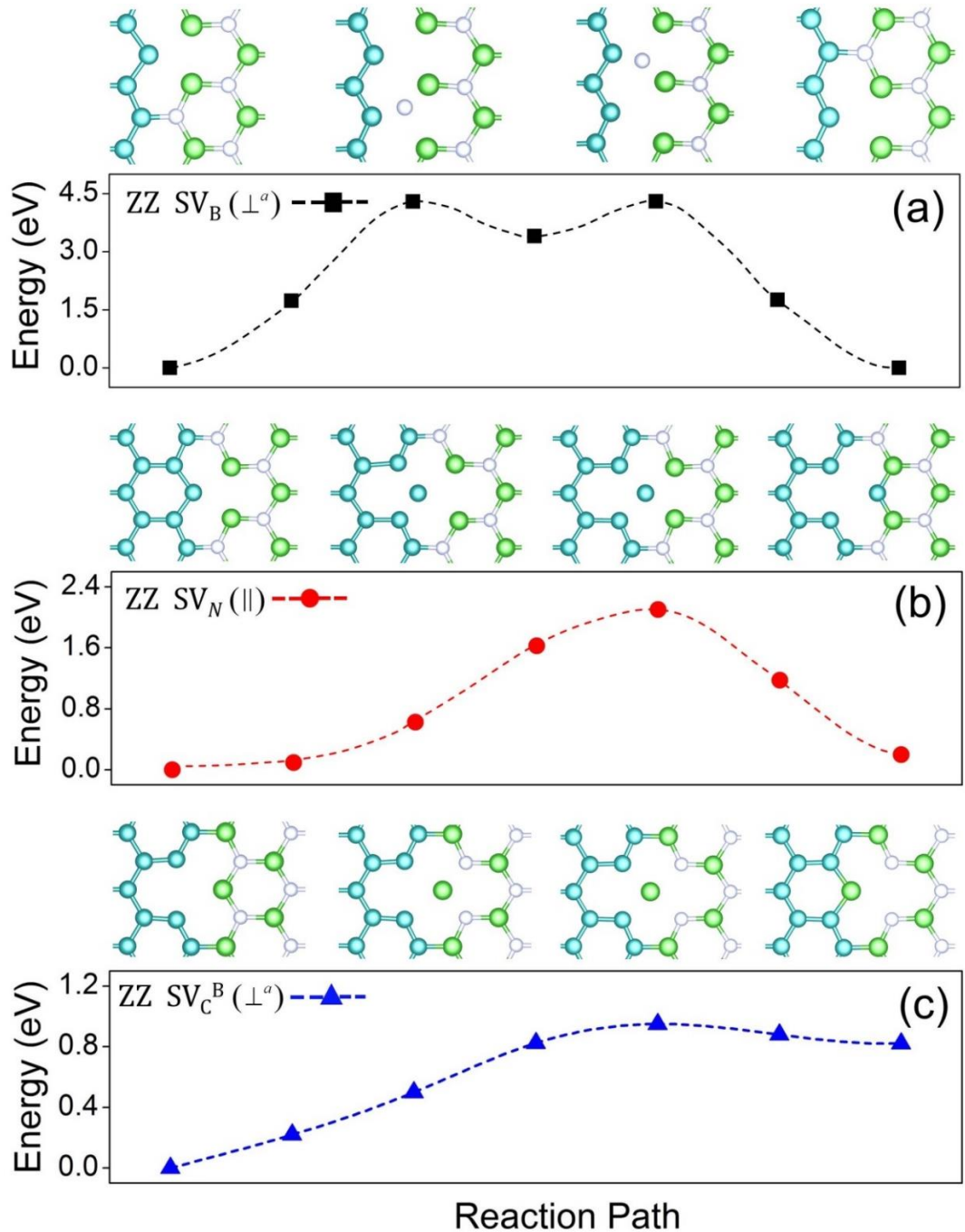

**Figure 5:** Three representative MEP curves to illustrate the migration process of monovacancies, being *a)* the $\perp^a$ motion of $SV_B$ with vacancy type evolving as B→C, *b)* the $\parallel$ motion of $SV_N$ with vacancy type evolving as N→N, and *c)* the $\perp^a$ motion of $SV_C^B$ with vacancy type evolving as C→B, at the ZZ interface. The solid symbols indicate the data obtained from the NEB calculations while the dashlines are used to guide the eye. For each case, several atomic configurations of the vacancy at different stages of the migration are presented to illustrate the evolution of local vacancy geometry, where atoms are colored as follows: C (dark cyan), B (green) and N (white).

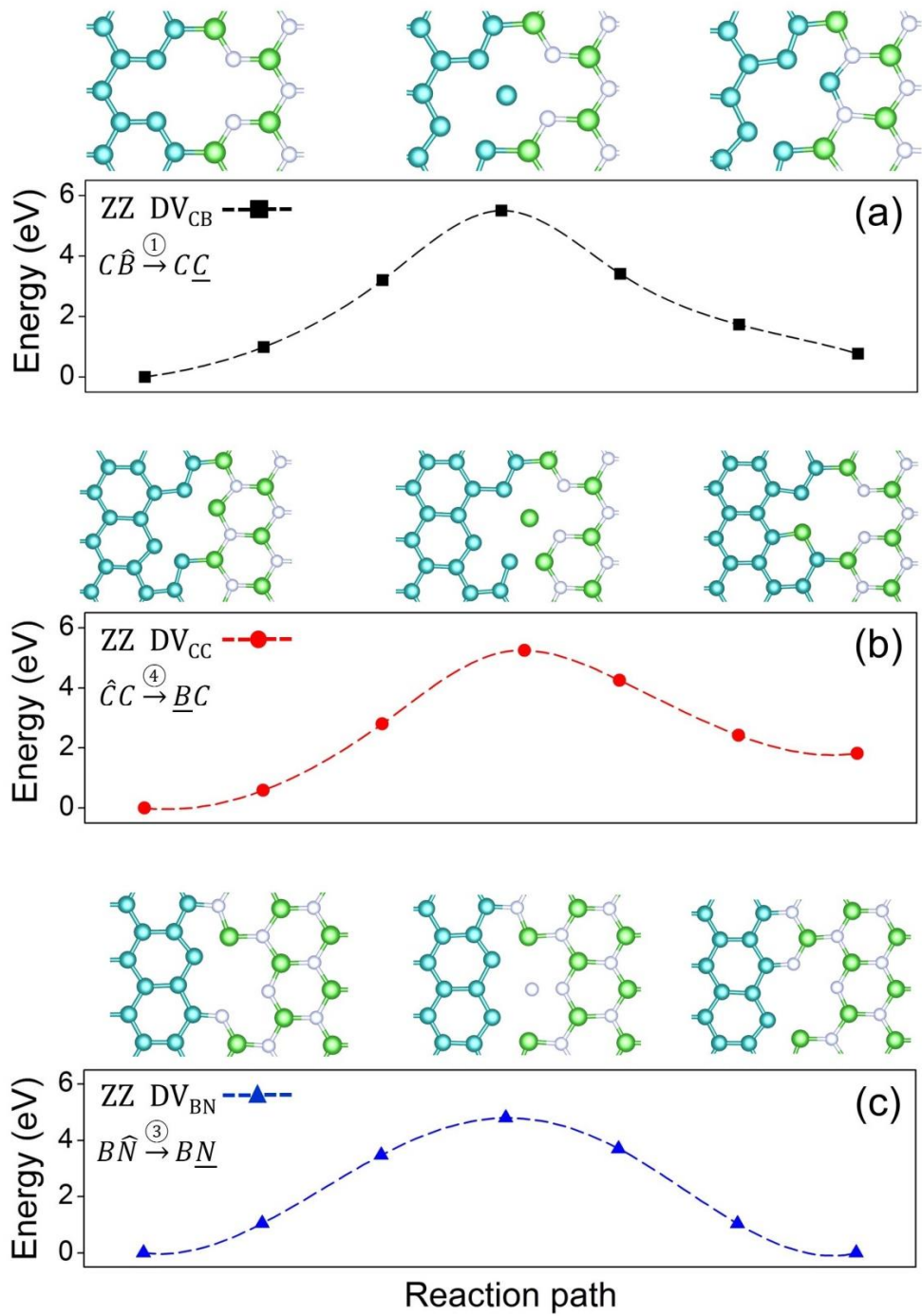

**Figure 6**: Three representative MEP curves to illustrate the migration process of divacancies, being *a*) a DV$_{CB}$ rotating via the motion of $C\hat{B} \overset{①}{\rightarrow} C\underline{C}$, *b*) a DV$_{CC}$ rotating via the motion of $\hat{C}C \overset{④}{\rightarrow} \underline{B}C$, *c*) a DV$_{BN}$ rotating via the motion of $B\hat{N} \overset{③}{\rightarrow} B\underline{N}$, at the ZZ interface (cf. Figure 4 and Table V). The solid symbols indicate the data obtained from the NEB calculations while the dashlines are used to guide the eye. For each case, several atomic configurations of the divacancy at different stages of the migration are presented to illustrate the evolution of local vacancy geometry, where different atoms are colored as follows: C (dark cyan), B (green) and N (white).